\documentclass[10pt,aps,prx,twocolumn,superscriptaddress,nofootinbib,floatfix]{revtex4-2}

\usepackage[english]{babel}

\usepackage{amsmath,amsfonts}
\usepackage{graphicx}
\usepackage[colorlinks=true, allcolors=blue]{hyperref}
\usepackage[table, dvipsnames]{xcolor} 
\usepackage{comment}

\usepackage{subcaption}
\usepackage{cleveref}
\usepackage{dsfont}
\usepackage{multirow}
\usepackage{booktabs}
\usepackage{orcidlink}
\usepackage[normalem]{ulem}
\usepackage{ragged2e}

\newcommand{\ket}[1]{\left|#1\right\rangle}
\newcommand{\bra}[1]{\left\langle#1\right|}

\begin{document}

\title{Quantum Reservoir Computing with Neutral Atoms on a Small, Complex, Medical Dataset}

\affiliation{Centre for Quantum Information, Simulation and Algorithms, The University of Western Australia, Perth, Australia}
\affiliation{Centre for Respiratory Health, University of Western Australia, Perth, WA, Australia}
\affiliation{School of Medicine $\&$ Pharmacology, University of Western Australia, Perth, Western Australia, Australia}
\affiliation{Dept of Respiratory Medicine, Fiona Stanley Hospital, Perth, WA, Australia}
\affiliation{Pawsey Supercomputing Research Centre, Perth, WA, 6152, Australia}
\affiliation{QuEra Computing Inc., 1284 Soldiers Field Road, Boston, MA, 02135, USA}
\affiliation{School of Physics, Mathematics and Computing, The University of Western Australia, 35 Stirling Hwy, Crawley WA, 6009, Australia}
\affiliation{These authors jointly supervised this work}

\author{Luke Antoncich\orcidlink{0009-0004-9124-2650}} 
\affiliation{Centre for Quantum Information, Simulation and Algorithms, The University of Western Australia, Perth, Australia}

\author{Yuben Moodley\orcidlink{0000-0002-0777-1196}}
\affiliation{Centre for Respiratory Health, University of Western Australia, Perth, WA, Australia}
\affiliation{School of Medicine $\&$ Pharmacology, University of Western Australia, Perth, Western Australia, Australia}
\affiliation{Dept of Respiratory Medicine, Fiona Stanley Hospital, Perth, WA, Australia}

\author{Ugo Varetto}
\affiliation{Pawsey Supercomputing Research Centre, Perth, WA, 6152, Australia}
\affiliation{Centre for Quantum Information, Simulation and Algorithms, The University of Western Australia, Perth, Australia}

\author{Jingbo Wang\orcidlink{0000-0001-7544-0084}}
\affiliation{Centre for Quantum Information, Simulation and Algorithms, The University of Western Australia, Perth, Australia}

\author{Jonathan Wurtz \orcidlink{0000-0001-7237-0789}}
\affiliation{QuEra Computing Inc., 1284 Soldiers Field Road, Boston, MA, 02135, USA}

\author{Jing Chen \orcidlink{0000-0003-0538-689X}}
\affiliation{QuEra Computing Inc., 1284 Soldiers Field Road, Boston, MA, 02135, USA}

\author{Pascal Jahan Elahi\orcidlink{0000-0002-6154-7224}}
\affiliation{Pawsey Supercomputing Research Centre, Perth, WA, 6152, Australia}
\affiliation{Centre for Quantum Information, Simulation and Algorithms, The University of Western Australia, Perth, Australia}
\affiliation{These authors jointly supervised this work}

\author{Casey R.~Myers\orcidlink{0000-0002-8838-7523}} 
\affiliation{School of Physics, Mathematics and Computing, The University of Western Australia, 35 Stirling Hwy, Crawley WA, 6009, Australia}
\affiliation{Pawsey Supercomputing Research Centre, Perth, WA, 6152, Australia}
\affiliation{These authors jointly supervised this work}



\begin{abstract}
Biomarker-based prediction of clinical outcomes is challenging due to nonlinear relationships, correlated features, and the limited size of many medical datasets. Classical machine-learning methods can struggle under these conditions, motivating the search for alternatives. In this work, we investigate quantum reservoir computing (QRC), using both noiseless emulation and hardware execution on the neutral-atom Rydberg processor \textit{Aquila}. We evaluate performance with six classical machine-learning models and use SHAP to generate feature subsets. We find that models trained on emulated quantum features achieve mean test accuracies comparable to those trained on classical features, but have higher training accuracies and greater variability over data splits, consistent with overfitting. When comparing hardware execution of QRC to noiseless emulation, the models are more robust over different data splits and often exhibit statistically significant improvements in mean test accuracy. This combination of improved accuracy and increased stability is suggestive of a regularising effect induced by hardware execution. To investigate the origin of this behaviour, we examine the statistical differences between hardware and emulated quantum feature distributions. We find that hardware execution applies a structured, time-dependent transformation characterised by compression toward the mean and a progressive reduction in mutual information relative to emulation.
\end{abstract}

\maketitle

\section{Introduction}

Biomarkers are measurable characteristics of an organism that can be used for diagnosis and prognosis. Their relationship with clinical outcomes is often nonlinear and involves correlated features. This limits the effectiveness of linear models and standard statistical methods, particularly in the small-data regime common in clinical studies \cite{ng_benefits_2023, ghassemi2020review}. Classical machine-learning methods can partially address these challenges by approximating nonlinear mappings and exploiting feature dependencies \cite{mi2021permutation, diaz2022ten, bavikadi2025machine, dai2025high}. However, high-capacity models are often prone to overfitting, while overly restrictive models may fail to capture relevant structure in the data \cite{HastieBook2009}.

This motivates the search for alternative techniques. A natural avenue of exploration in this domain is quantum machine-learning (QML)~\cite{Biamonte_2017}. Quantum dynamics are known to generate statistical patterns and correlations that cannot, in general, be efficiently reproduced by classical computers. Quantum processes may be well suited to transforming classical data in a way that exposes nonlinear structure inaccessible to purely classical transformations~\cite{Goto_2021}.

Quantum reservoir computing (QRC) provides a concrete framework for exploiting this idea. Inspired by classical reservoir computing~\cite{jaeger-haas, maass}, QRC uses the fixed time evolution of a quantum many-body system as a nonlinear feature map \cite{fuuji-nakajima}. Classical input data are encoded into the quantum system, which evolves according to its Hamiltonian. The expectation values of selected observables at multiple evolution times form the output quantum feature vector. A classical readout model is trained on this output, while the quantum dynamics themselves remain untrained. In the framework of neural networks, the evolution of the quantum system could be considered a hidden layer of neurons, mapping the input layer to the output layer.

QRC has been explored in a range of application domains, including image denoising \cite{das2025imagedenoisingquantumreservoir}, economic forecasting \cite{li2025quantumreservoircomputingrealized, domingo2023optimalquantumreservoircomputing, vitali2025quantumreservoircomputingcredit}, and materials science \cite{xin2025unsuperviseddetectiontopologicalphase, tandon2025quantumreservoircomputingcorrosion}. Relative to other quantum machine-learning paradigms, QRC avoids the need for gradient-based optimisation over large quantum state spaces, thereby sidestepping issues such as barren plateaus \cite{mcclean2018barren}. QRC can also be implemented on near-term quantum hardware without requiring fault tolerance \cite{kornjaca_large-scale_2024, dudas2023quantum, cimini2025largescalequantumreservoircomputing}. Recent work has suggested that QRC-based approaches may retain robustness as dataset size decreases, making them appealing for biomedical settings \cite{beaulieu_robust_2024}.

In this study, we investigate the performance of QRC on a binary classification task with biomarker data. Using six classical machine-learning models, we first establish a performance baseline using the entire dataset. We then use SHAP-based feature ranking to identify informative subsets of biomarkers. These feature subsets are encoded into an emulated quantum reservoir to generate corresponding quantum feature sets. Downstream machine-learning performance is then evaluated using both the classical data and the emulated quantum data.

We find that, when evaluated at their respective optimal feature counts, the emulated QRC approach achieves mean test accuracies comparable to the classical approach. However, models trained on emulated quantum features typically exhibit higher training accuracy and increased split-to-split variability, consistent with mild overfitting. This indicates that noiseless emulation of the quantum reservoir is not transforming the data in a way that provides an advantage for downstream learning.

We also test hardware implementation of QRC, using the neutral-atom quantum processor \textit{Aquila}. Due to restrictions on available hardware shots, the biomarker subsets are aggregated over multiple splits of the data, rather than being unique per split. We compare this hardware implementation to emulated QRC with identical biomarker subsets. We observe that machine-learning models trained on hardware features are often more robust over different data splits and/or have higher mean test accuracy. Further analysis reveals these results are statistically significant across multiple model and feature-count configurations.

To better understand the origin of this discrepancy between hardware and emulation, we analyse the quantum features at the level of individual expectation values. We find that hardware execution applies a structured, time-dependent transformation to these feature distributions, characterised by compression toward the mean and a progressive reduction in mutual information between hardware and emulated features. These observations suggest that hardware-specific effects systematically reshape the quantum feature representation in a manner consistent with a regularising transformation.

The remainder of this paper is organised as follows. Section~\ref{methods} describes the biomarker dataset, classical and quantum feature-generation pipelines, and the evaluation framework. Section~\ref{results} presents the comparative performance of classical, emulated QRC, and hardware QRC approaches, together with an analysis of the statistical differences between hardware and emulated quantum feature representations. Section~\ref{conclusion} summarises the main findings and outlines directions for future work.

\section{Materials \& Methods}
\label{methods}

\subsection{Dataset and Preprocessing}

The dataset used in this work was provided by the Centre for Respiratory Health in Western Australia. After cleaning, the dataset consisted of 108 samples each with 56 features. Most features correspond to either biomarkers measured at baseline, or their changes over a three-month period. Additional features include demographic information such as age and gender. The predictive task is a binary classification: determining whether a patient will experience a decline in function one year after the initial measurement. The learning task is balanced; there are an equal number of patients in each class.

There were a small number of missing values which were imputed using the mean value of the corresponding feature. Subsequently, all features were scaled to the range [0, 1] using Min–Max normalisation, which is required for stable training and encoding into the quantum reservoir.

\subsection{Machine-Learning Models}

We evaluated six supervised learning algorithms widely used for tabular prediction tasks:

\paragraph{XGBoost} \cite{xgboost}:
An efficient implementation of gradient boosting that incorporates regularisation, weighted tree growth, and optimised handling of sparse features. XGBoost is typically strong on small-to-medium tabular datasets and serves as a strong classical baseline.

\paragraph{Random Forest} \cite{breiman_random_2001}:
An ensemble of decision trees trained on bootstrap samples of the data. Each tree uses a random subset of features at each split, reducing correlation between trees and improving robustness.

\paragraph{Gradient Boosted Trees} \cite{gradientboost,FRIEDMAN2002367}:
An additive ensemble model that iteratively fits small decision trees, each correcting the residual errors of the previous ones. This sequential structure allows the method to capture complex feature interactions.

\paragraph{Logistic Regression} \cite{hosmer2013logistic}:
A linear classifier that models the log-odds of the binary outcome as a weighted combination of the input features. Regularisation is applied to prevent overfitting.

\paragraph{Support Vector Machine (SVM)} \cite{cortes_support-vector_1995}:
A margin-based classifier that identifies the separating boundary between classes. Both linear and radial-basis function (RBF) kernels were considered, allowing either a linear or non-linear decision surface.

\paragraph{k-Nearest Neighbours (kNN)} \cite{kNN}:
A non-parametric classifier that assigns labels based on the majority class among the \(k\) closest training samples in feature space. The method is sensitive to feature scaling and local data structure.

All models were implemented using \texttt{scikit-learn} \cite{pedregosa2011scikit}, except for XGBoost which was implemented using the \texttt{xgboost} package.

\subsection{Machine-Learning Hyperparameter Optimisation}

All classical prediction models considered in this work require the specification of \emph{machine-learning hyperparameters}, such as regularisation strengths, tree depths, learning rates, kernel parameters, or the number of estimators. These hyperparameters control the inductive bias and optimisation behaviour of the learning algorithm and are distinct from the \emph{quantum parameters} of the reservoir (e.g.\ Rabi frequencies, detunings, or evolution times), which define the physical generation of the quantum features. In this study, quantum reservoir parameters were fixed, while all machine-learning hyperparameters were optimised in a data-driven manner.

To obtain unbiased estimates of generalisation performance, hyperparameter optimisation was performed using a nested cross-validation framework. The full dataset was first partitioned into five independent outer splits, each consisting of an 80/20 train-test division. The held-out test sets in the outer loop were reserved exclusively for final performance evaluation.

Within each outer training set $\mathcal{D}_{\text{train}}$, hyperparameter tuning was carried out in an inner loop using stratified $K$-fold cross-validation with $K=10$, preserving class proportions across folds \cite{kohavi_study_1995, hastie_elements_2009}. The training set was decomposed as
\[
\mathcal{D}_{\text{train}} = \bigcup_{k=1}^{K} \mathcal{D}_{\text{train}}^{(k)},
\]
where each fold $\mathcal{D}_{\text{train}}^{(k)}$ is disjoint and of approximately equal size. For a given hyperparameter configuration, models were trained on $\mathcal{D}_{\text{train}} \setminus \mathcal{D}_{\text{train}}^{(k)}$ and evaluated on $\mathcal{D}_{\text{train}}^{(k)}$, yielding validation accuracies $a^{(k)}$. The cross-validation score for that configuration was defined as the mean validation accuracy,
\[
\bar{a} = \frac{1}{K} \sum_{k=1}^{K} a^{(k)}.
\]

Hyperparameters were selected using random search \cite{bergstra_random_2012}. For each learning algorithm, probability distributions were specified over all tunable hyperparameters (Appendix~\ref{app:hyperparams}), and $M=1000$ configurations were sampled independently from the resulting joint distribution. Each configuration was evaluated using the inner-loop cross-validation procedure described above, and the configuration maximising $\bar{a}$ was selected. Fixed random seeds were used throughout to ensure reproducibility of data splits and hyperparameter sampling.

This optimisation strategy was applied consistently in three settings. First, it was used to obtain a single optimally tuned model per algorithm and outer split when training on the full set of classical input features. These models define the baseline classical performance and were subsequently used for SHAP-based feature ranking. Second, following SHAP analysis, models were retrained using nested subsets of the top-$k$ ranked features. For each feature count $k$, a separate hyperparameter optimisation was performed within the corresponding outer training split, using the same random search samples to ensure that performance differences across $k$ reflect changes in input dimensionality rather than stochastic variation in the search procedure. Third, the identical hyperparameter optimisation protocol was applied when training classical models on quantum feature vectors generated by the quantum reservoir, for both emulated and hardware implementations.

In all cases, the final performance reported for a given model and feature count corresponds to evaluation on the held-out test sets of each outer split, ensuring a fair and consistent comparison between classical features and quantum-generated features across varying input dimensionalities.

\subsection{SHAP Analysis and Feature Ranking}

To obtain estimates of feature importance for each model, we applied SHAP (SHapley Additive exPlanations)~\cite{shap}, a framework based on Shapley values from cooperative game theory. SHAP attributes a model’s prediction to individual input features by quantifying how each feature contributes to the difference between the model output and its expected value.

For a model $f$ and input $\mathbf{x}$, the Shapley value $\phi_i$ assigned to feature $i$ is defined as
\begin{equation*}
\phi_i = \sum_{S \subseteq N \setminus \{i\}} 
\frac{|S|! \, (|N| - |S| - 1)!}{|N|!} 
\left[f_{S \cup \{i\}}(\mathbf{x}) - f_S(\mathbf{x})\right],
\end{equation*}
where $N$ is the set of all features, and $f_S(\mathbf{x})$ denotes the model output when only the features in subset $S$ are considered. Shapley values satisfy the axioms of Efficiency, Symmetry, Linearity, and Null player~\cite{Shapley+1953+307+318}, making them a principled and model-agnostic approach to feature attribution.

For any instance $\mathbf{x}$, SHAP provides an additive decomposition of the model prediction,
\begin{equation*}
f(\mathbf{x}) = 
\mathbb{E}[f(\mathbf{x})] + 
\sum_{i=1}^{|N|} \phi_i,
\end{equation*}
ensuring that the sum of feature contributions equals the deviation of the prediction from the dataset mean. To obtain a measure of feature importance for a given model instance (i.e., a fixed algorithm and outer split), we computed the absolute Shapley value for each feature using a model trained on the entire set of training data. Features were then ranked in descending order of their absolute SHAP values.

These rankings were used to construct the nested feature subsets employed in subsequent classical and emulated QRC experiments, with feature selection performed independently for each algorithm and outer split. A visual comparison of the ten most influential features identified by each model, averaged over all outer splits, is provided in Appendix~\ref{app:initial_shap} (Figure~\ref{fig:shap_top10}).

\subsection{Quantum Reservoir Encoding}

The quantum reservoir was implemented using both numerical emulation on Bloqade~\cite{bloqade} and hardware execution on \textit{Aquila}, a programmable neutral-atom quantum processor developed by QuEra Computing~\cite{wurtz_aquila_2023}. \textit{Aquila} employs arrays of $^{87}$Rb atoms trapped in optical tweezers and excited to high-lying Rydberg states, enabling analogue realisation of interacting spin systems with tunable global and local controls.

The quantum reservoir dynamics are governed by the time-dependent Rydberg Hamiltonian
\cite{wurtz_aquila_2023}:
\begin{align}
H(t)=&\frac{\Omega(t)}{2}\sum_i 
\left( e^{i\phi(t)}\ket{g_i}\bra{r_i} + e^{-i\phi(t)}\ket{r_i}\bra{g_i} \right) \notag\\
&- \sum_i \left( \Delta_{g}(t) + \alpha_i \Delta_{l}(t) \right) \hat{n}_i \notag\\
&+ \sum_{i<j}\frac{C_6}{|\vec{x}_i-\vec{x}_j|^6}\hat{n}_i \hat{n}_j,
\label{rydberg-hamiltonian}
\end{align}
where $\ket{g}$ and $\ket{r}$ denote the ground and Rydberg states, $\hat{n}_i=\ket{r_i}\bra{r_i}$ is the Rydberg occupation operator, $\Omega(t)$ is the global Rabi drive with phase $\phi(t)$, $\Delta_g(t)$ is the global detuning, $\Delta_l(t)$ is a local detuning term modulated by coefficients $\alpha_i$, and the interaction strength is set by fixed atom positions $\vec{x}_i$ with $C_6 = 5{,}420{,}503\,\mu\mathrm{m}^6\,\mathrm{rad}/\mu\mathrm{s}$.

Classical input data are encoded into the quantum reservoir through the local detuning coefficients $\alpha_i$. For a given feature subset of size $k$, the $k$ SHAP-ranked features are assigned to $k$ atoms arranged in a one-dimensional chain with open boundary conditions. Each atom encodes a single data feature, ordered by decreasing SHAP importance. From this point onward, we use the terms \emph{number of data features} and \emph{number of atoms} interchangeably, referring to the dimensionality of the classical input and the size of the quantum reservoir, respectively.

The detuning applied to each atom consists of a fixed global component and a feature-dependent local contribution. The global detuning was set to
\[
\Delta_g = 4.5\;\mathrm{rad}/\mu\mathrm{s},
\]
while the local detuning scale was fixed to
\[
\Delta_l = -9.0\;\mathrm{rad}/\mu\mathrm{s}.
\]
The resulting effective detuning on atom~$i$ is therefore
\[
\Delta_i = \Delta_g + \alpha_i \Delta_l.
\]
This protocol was chosen such that the effective detuning at each atom is spread over a range of both positive and negative values. For a positive detuning, an atom will generally find it energetically favourable to be in the Rydberg state. The opposite holds for a negative detuning. As such, the value of the encoded data feature can have a distinctly different impact on the time evolution of the atom which encodes it. The interactions between nearby atoms can then reveal correlations between the underlying data features.

A second reason for this protocol is that when running on \textit{Aquila}, detuning applied globally yields more accurate and stable dynamics than detuning implemented locally at individual atoms. This is the case even when encoding identical target detuning values. However, global detuning on the hardware is restricted to non-negative values. This motivates the hybrid scheme, which enables a broader range of effective detuning values to be represented while preserving the stability provided by a strong global offset.

All remaining Hamiltonian parameters were held fixed across experiments. The global Rabi frequency was set to a comparatively large value,
\[
\Omega = 2\pi\;\mathrm{rad}/\mu\mathrm{s},
\]
allowing the reservoir dynamics to develop rapidly and enabling the full time evolution to be completed within the available coherence window. The drive phase was fixed to $\phi = 0$, ensuring a common rotation axis for all atoms. Interatomic separations were fixed at
\[
|\vec{x}_i - \vec{x}_j| = 10\;\mu\mathrm{m}.
\]
This spacing was chosen to balance interaction strength and reservoir expressivity: atoms placed too closely become strongly blockaded into a restricted subspace, while atoms placed too far apart experience interactions that are too weak to generate sufficient entanglement over the evolution time.

On hardware, atom positions are constrained to a finite trapping region of approximately $125 \times 75~\mu\mathrm{m}$. Given the fixed interatomic spacing of $10~\mu\mathrm{m}$ and the use of a one-dimensional chain geometry, this available window limits the maximum reservoir size that can be realised. For this reason, we examined reservoirs containing up to a maximum of 14 atoms, ensuring that all configurations fit within the hardware bounds while maintaining uniform spacing and interaction strength across the array.

The reservoir state was evolved under this Hamiltonian for five discrete time steps over a total duration of $4.0\;\mu\mathrm{s}$. This evolution time was chosen to allow sufficient nonlinear many-body dynamics to develop while remaining short enough to maximise coherence. The readout consists of expectation values of single-site and two-site observables, specifically $\langle Z_i \rangle$ and $\langle Z_i Z_j \rangle$. We evaluate these expectation values at multiple time steps, and then concatenate to form a high-dimensional quantum feature vector.

\subsection{Shot-Efficient Hardware Feature Selection}

Executing the quantum reservoir encoding on hardware requires repeated experimental runs (“shots”) to estimate each expectation value, with statistical uncertainty decreasing only as the inverse square root of the number of shots. Because shots on neutral-atom platforms are a limited and costly resource, directly evaluating every learning algorithm, feature count, and SHAP ranking arising from the outer cross-validation splits would require a large number of shots.

To mitigate this constraint while retaining representative feature structure, we examined the mean SHAP rankings obtained during the initial classical feature-selection stage (Appendix~\ref{app:initial_shap}). These analyses revealed two rough groupings: the three tree-based models (XGBoost, Random Forest, and Gradient Boosted Trees) produced similar feature rankings, while Logistic Regression, SVM, and kNN formed a second group with a distinct but likewise consistent ordering. Based on this observation, we constructed two aggregated SHAP-ranked feature sets and performed hardware encoding only for these representative orderings, reducing the required shot count by a factor of fifteen.

Aggregating SHAP rankings across the five outer splits does introduce the possibility of information leakage from held-out test partitions into the aggregated orderings. If hardware shot cost had permitted, we would have evaluated all split-specific feature subsets independently. Given current experimental constraints, the aggregated approach represents a practical compromise that preserves the dominant feature-selection structure while enabling feasible hardware execution.

\subsection{Model Evaluation and Comparison}

All final model evaluations were performed on the held-out 20\% test sets associated with each of the five outer data splits.

The primary evaluation metric was test accuracy, averaged over the five outer splits. We compare classical and QRC performance at matched feature counts~$k$, enabling an assessment of how accuracy scales as additional clinically meaningful features are incorporated, and whether quantum features provide an advantage over classical features at equivalent levels of input information.

To quantify the relative performance between classical features and quantum features, we additionally computed a standardised difference for each matched feature count~$k$. This metric measures the difference in mean accuracy between two models relative to the uncertainty associated with that difference, which provides an indication of how confidently an observed performance gap can be attributed to a genuine effect rather than to split-to-split variability.

Let $\mu_{\mathrm{qrc}}$ and $\mu_{\mathrm{class}}$ denote the mean test accuracies for QRC and classical models, with standard deviations $\sigma_{\mathrm{qrc}}$ and $\sigma_{\mathrm{class}}$ computed across $n_{\mathrm{qrc}}$ and $n_{\mathrm{class}}$ outer splits, respectively. The standard error of the difference is
\[
\mathrm{SE}_{\mathrm{diff}}
  = \sqrt{
      \left( \frac{\sigma_{\mathrm{qrc}}}{\sqrt{n_{\mathrm{qrc}}}} \right)^{\!2}
      +
      \left( \frac{\sigma_{\mathrm{class}}}{\sqrt{n_{\mathrm{class}}}} \right)^{\!2}
    }.
\]
The standardised difference is then defined as
\begin{align}
\Delta_{\mathrm{std}}
   = \frac{ \mu_{\mathrm{qrc}} - \mu_{\mathrm{class}} }{ \mathrm{SE}_{\mathrm{diff}} }.
\label{Eq:standardiseddiff}
\end{align}

This quantity is analogous to a signal-to-noise ratio or a $z$-score: values of $|\Delta_{\mathrm{std}}| \lesssim 1$ indicate that observed differences are comparable to the variability across splits and should therefore be interpreted cautiously, while larger magnitudes suggest increasing confidence that the observed performance difference reflects a reproducible effect. Plotting $\Delta_{\mathrm{std}}$ across feature counts thus provides a scale-independent view of both the direction and robustness of performance differences, complementing mean accuracy comparisons in the presence of substantial dataset variability.

\begin{figure*}[ht]
    \centering
    \includegraphics[width=\linewidth]{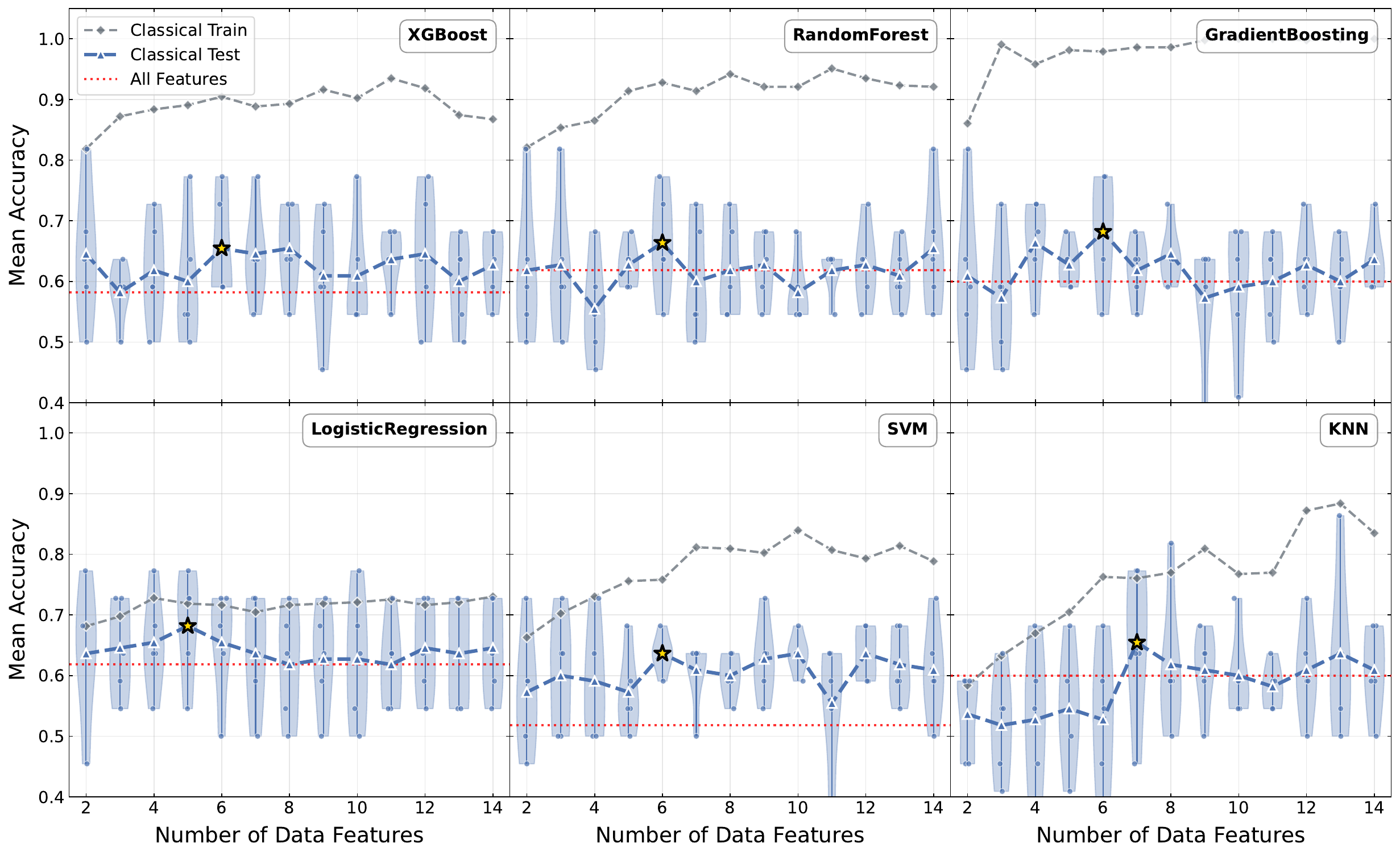}
    \caption{\justifying Performance of six machine-learning models trained on SHAP-ranked classical feature subsets. We use five unique train/test splits of the data. For each model, data split, and feature count, we evaluate performance with optimal machine-learning hyperparameters. Violin plots show variability in test accuracy across the five data splits. The highest mean test accuracy is marked by the black star. The dotted line marks the mean test accuracy when using all 56 data features. In all cases the models exhibit improved performance when using a subset of all available data features.}
    \label{fig:classical_performance}
\end{figure*}

\section{Results and Discussion}
\label{results}

\subsection{Model Performance}
\label{model_performance}

\subsubsection{With Classical Data Features}
\label{results_classical}

\begin{figure*}[ht]
    \centering
    \includegraphics[width=\linewidth]{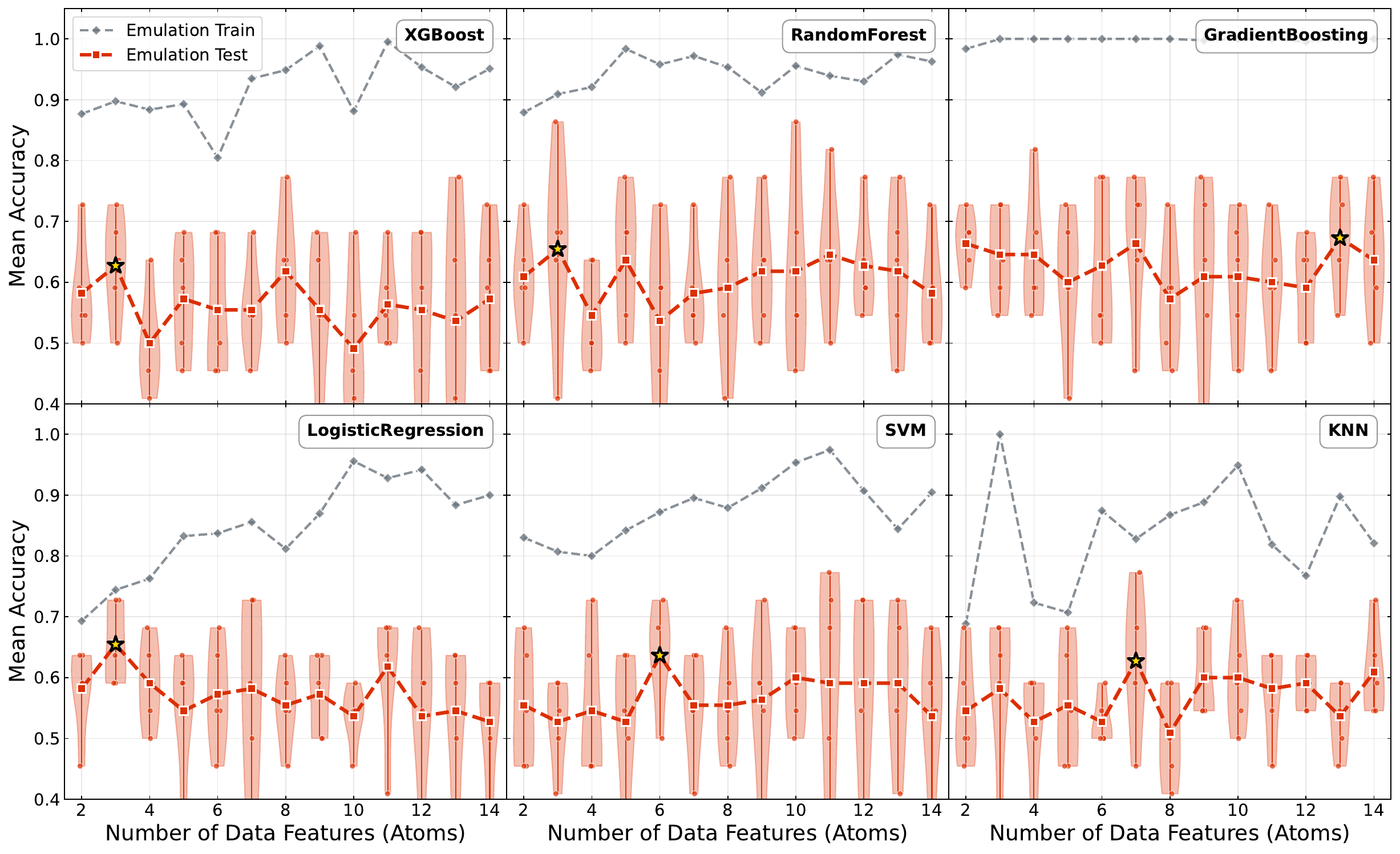}
    \caption{\justifying Performance of six machine-learning models trained on quantum features generated by an emulated quantum reservoir. The same SHAP-ranked feature subsets as in the classical results are used to encode data into the atoms of the reservoir. We use five unique train/test splits of the data. For each model, data split, and atom count, we evaluate performance with optimal machine-learning hyperparameters. Violin plots show variability in test accuracy across the five data splits. The highest mean test accuracy is marked by the black star. The optimal performance achieved using emulated quantum features is comparable to that obtained with classical features. Models using emulated quantum features generally have higher training accuracies and larger variability across data splits, indicative of overfitting.}
    \label{fig:qrc_emulated_performance}
\end{figure*}

Figure~\ref{fig:classical_performance} shows the performance of the six machine-learning models as the number of classical data features increases from 2 to 14. Each violin plot summarises the distribution of test accuracies across the five outer 80/20 data splits, while the black star denotes the highest mean test accuracy for each model. The dashed red line indicates the performance obtained when using the full set of available data features.

Across all models, the optimal test performance occurs within a narrow range of feature counts, between five and seven features. This optimal point always exceeds the performance obtained when using all available features, often by a substantial margin. In many cases the performance achieved across a broad range of feature subsets is comparable to, or better than, that obtained with the full feature set. This is a solid justification for using this feature reduction method on this dataset.

One exception is kNN with low feature counts. For kNN, reducing the number of input features below its optimal value of seven leads to a sharp degradation in performance, with accuracies approaching random guessing at low feature counts.

A closer inspection of the kNN results reveals that this sharp performance decrease coincides with the removal of \texttt{Feature~2} from several of the outer splits (see Appendix~\ref{app:initial_shap}). This feature was identified as the most important feature by all five other models, yet it appears lower in the SHAP ranking for kNN. Despite this, including this feature in the kNN input leads to a pronounced improvement in test accuracy, indicating that it carries strong discriminative information for distance-based classification once present in the feature space.

A more in-depth analysis of why SHAP did not identify \texttt{Feature~2} as highly important for kNN is beyond the scope of this work. However, this observation is consistent with broader discussions in the interpretability literature noting that additive feature-attribution methods may not fully capture feature relevance in models whose predictions depend on global changes to the geometry of the feature space~\cite{molnar2021generalpitfallsmodelagnosticinterpretation, molnar_interpretable_2022}.

Across all models, substantial variability is observed across the five outer splits, as reflected by the height of the violin plots. This split-to-split variability indicates that performance is highly sensitive to the specific data partition, a common characteristic of small biomedical datasets with limited sample sizes.

\subsubsection{With Quantum Features from Emulation}
\label{results_emulation}

Figure~\ref{fig:qrc_emulated_performance} shows the performance of six machine-learning models trained on quantum reservoir features. The number of atoms ranges from 2 to 14 and the reservoir is emulated using Bloqade. As in the classical analysis, the violin plots summarise the distribution of test accuracies across the five outer dataset splits.

Across all six models, quantum features generated through emulation achieve mean test accuracies that are comparable to those obtained using classical features. In particular, the optimal test accuracy for each model is similar across both methods. No systematic improvement or degradation is observed, indicating that the emulated quantum reservoir is able to reproduce classical performance levels, rather than exceeding them.

Unlike the classical feature-selection results, there is no longer a consistently optimal input dimensionality across models. Several models achieve their highest mean test accuracy with as few as three atoms in the quantum reservoir, while Gradient Boosted Trees exhibit their optimal test accuracy with thirteen atoms. SVM and kNN achieve their best performance at six and seven atoms respectively, matching the optimal feature counts observed in the classical pipeline. Again, a spike in kNN performance coincides with the introduction of \texttt{Feature~2}.

A notable difference compared to the classical setting is that all models exhibit higher training accuracy when using emulated quantum features. For all models except kNN, this increase in training accuracy is accompanied by an increase in the variability of test accuracy across the five dataset partitions. Taken together, this pattern is consistent with what would be expected under increased overfitting to the training data.

If the performance of noiseless QRC in this domain is to be further improved, it would be necessary to understand the origin of these effects and to identify strategies that mitigate them. One natural class of approaches is regularisation, which explicitly constrains model complexity or suppresses sensitivity to fluctuations in the input features. Other complementary strategies may also be relevant, including dimensionality reduction of the quantum features, or modifications to the learning procedure that promote robustness in the small-data regime.

Overall, these results indicate that emulated quantum features can match the optimal performance of classical feature representations, while also exhibiting statistical signatures consistent with increased overfitting to the training data.

\subsection{Hardware Results}
\label{hardware_results}

\subsubsection{Controlled Hardware–Emulation Comparison}

Having established that classical features and quantum features from emulation exhibit comparable performance across a range of models and feature counts, we now turn to the analysis of hardware quantum features. The objective of this section is to assess whether experimental execution introduces qualitative differences when compared to noiseless emulation.

A direct comparison between hardware and classical baselines is not straightforward in this setting. The restriction on available hardware shots necessitates the use of aggregated SHAP-ranked feature subsets. Applying an equivalent aggregation procedure to classical features would introduce a heightened risk of test-set leakage, as the feature subsets are explicitly derived from classical model explanations. As a result, any apparent performance differences could not be cleanly attributed to the feature representations themselves.

To avoid this ambiguity, we restrict our analysis to a controlled comparison between hardware and emulation. Given that emulated and classical methods were shown in the preceding sections to achieve comparable performance, differences observed between hardware and emulation under these matched conditions serve as an indirect comparison between hardware and classical approaches.

\subsubsection{Model Performance}

\begin{figure}
    \centering
    \includegraphics[width=1\linewidth]{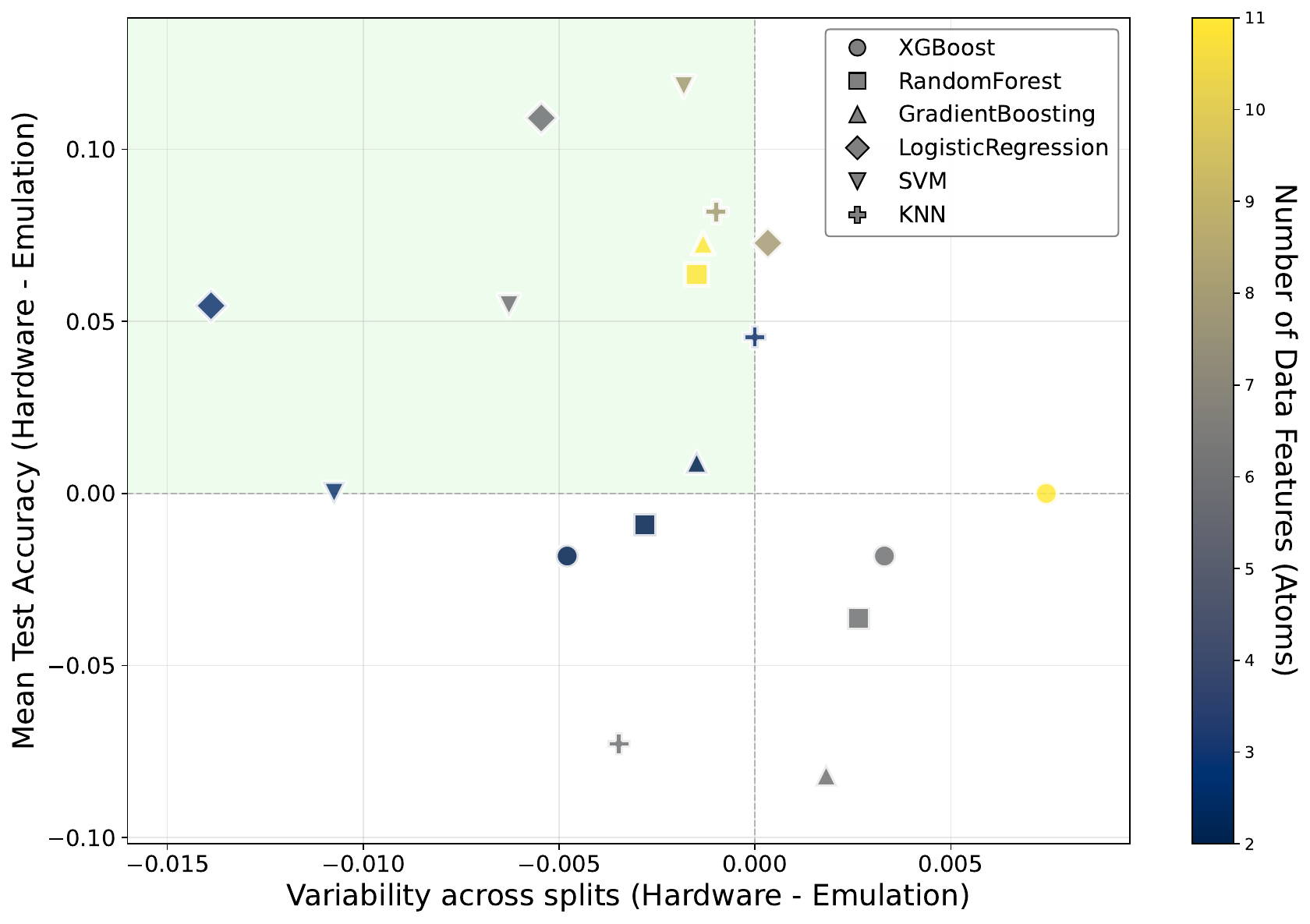}
    \caption{\justifying Changes in mean test accuracy (vertical axis) and split-to-split variability (horizontal axis) when moving from emulated to hardware quantum features. We observe that models trained on hardware quantum features often exhibit less variability across data splits than emulated quantum features. Importantly, this is often also accompanied by an increase in mean test accuracy. This indicates that executing the quantum reservoir on hardware helps to mitigate the overfitting tendencies observed with the emulated reservoir.}
    \label{fig:bias_variance}
\end{figure}

Figure~\ref{fig:bias_variance} summarises the effect of hardware execution on model performance for a restricted set of feature-count configurations. For each configuration, we report the change in mean test accuracy across outer splits on the vertical axis and the corresponding change in split-to-split variability on the horizontal axis, computed relative to emulated QRC.

The majority of configurations lie to the left of zero, indicating that hardware execution systematically reduces split-to-split variability compared to noiseless emulation. Crucially, this reduction in variability is not generally accompanied by a degradation in predictive performance. Instead, many configurations also exhibit improvements in mean test accuracy, resulting in a pronounced clustering of points in the upper-left quadrant of the plot.

It is also worth noting that of the four configurations appearing in the lower-right quadrant, corresponding to cases where hardware exhibits both increased variability and reduced test accuracy relative to emulation, three originate from the same hardware execution. This raises the possibility that these degraded outcomes reflect run-specific artefacts rather than a systematic effect, although confirming this would require repeated hardware trials.

When viewed in the context of Section~\ref{results_emulation}, these results suggest that the time evolution implemented by the hardware reservoir modifies the quantum features in a way that mitigates the overfitting tendencies observed under noiseless emulation. However, analysis of the training accuracies in Fig.~\ref{fig:hardware_violins} (Appendix~\ref{appendix:hardware_aggregate}) reveals that models trained on hardware features achieve training accuracies very similar to those obtained using emulated features. This indicates that hardware execution is not simply limiting model sensitivity to the training data. Instead, the improvement in generalisation appears to arise from a modification of the feature representation itself, which makes the resulting data more amenable to downstream learning.

\subsubsection{Standardised Difference Analysis}

\begin{figure}
    \centering
    \includegraphics[width=1\linewidth]{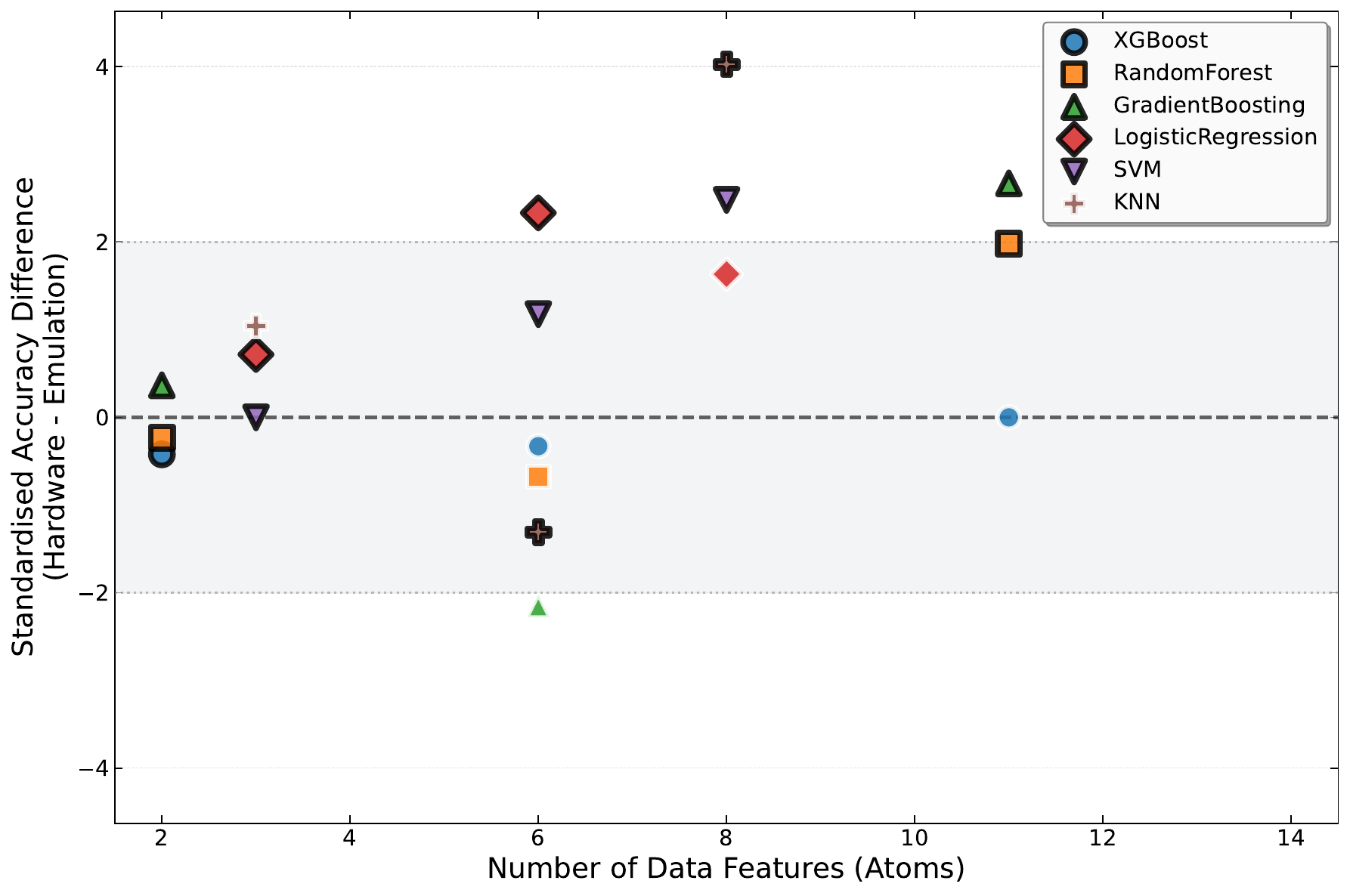}
    \caption{\justifying Standardised difference (see Eq.~\ref{Eq:standardiseddiff}) in mean test accuracy between models trained on hardware and emulated quantum features. Points with a black outline correspond to models which also exhibited reduced variability on hardware compared to emulation. Across most model and feature-count combinations, hardware performance is comparable to or exceeds that of emulation, with several configurations exhibiting statistically significant advantages for hardware execution (standardised difference $>2$). }
    \label{fig:standardised_difference_aggregate}
\end{figure}

Building on the performance trends observed in the previous section, we quantify differences between hardware and emulated quantum features using the \emph{standardised difference} in test accuracy for each model and feature-count combination. This metric compares the mean test accuracy of the two approaches while accounting for uncertainty across data splits, and is well suited to the substantial variability encountered in small-data settings.

Figure~\ref{fig:standardised_difference_aggregate} summarises the resulting comparison between hardware and emulation using identical feature encodings and learning pipelines. Across the majority of configurations, hardware performance is comparable to or exceeds that of emulation. In several cases, spanning multiple models and feature counts, hardware exhibits statistically significant improvements, with standardised differences exceeding a value of 2. In contrast, only a single configuration shows a statistically significant advantage for emulation. Certain models, such as Logistic Regression and SVM, consistently achieve higher performance when trained on hardware features.

Taken together, the statistically significant performance improvements observed here, alongside the reduced split-to-split variability reported in the previous section, motivate a deeper investigation of the differences between hardware and emulated quantum features.

\subsection{Comparison of Hardware and Emulated Quantum Features}

\begin{figure*}
    \centering
    \includegraphics[width=1\linewidth]{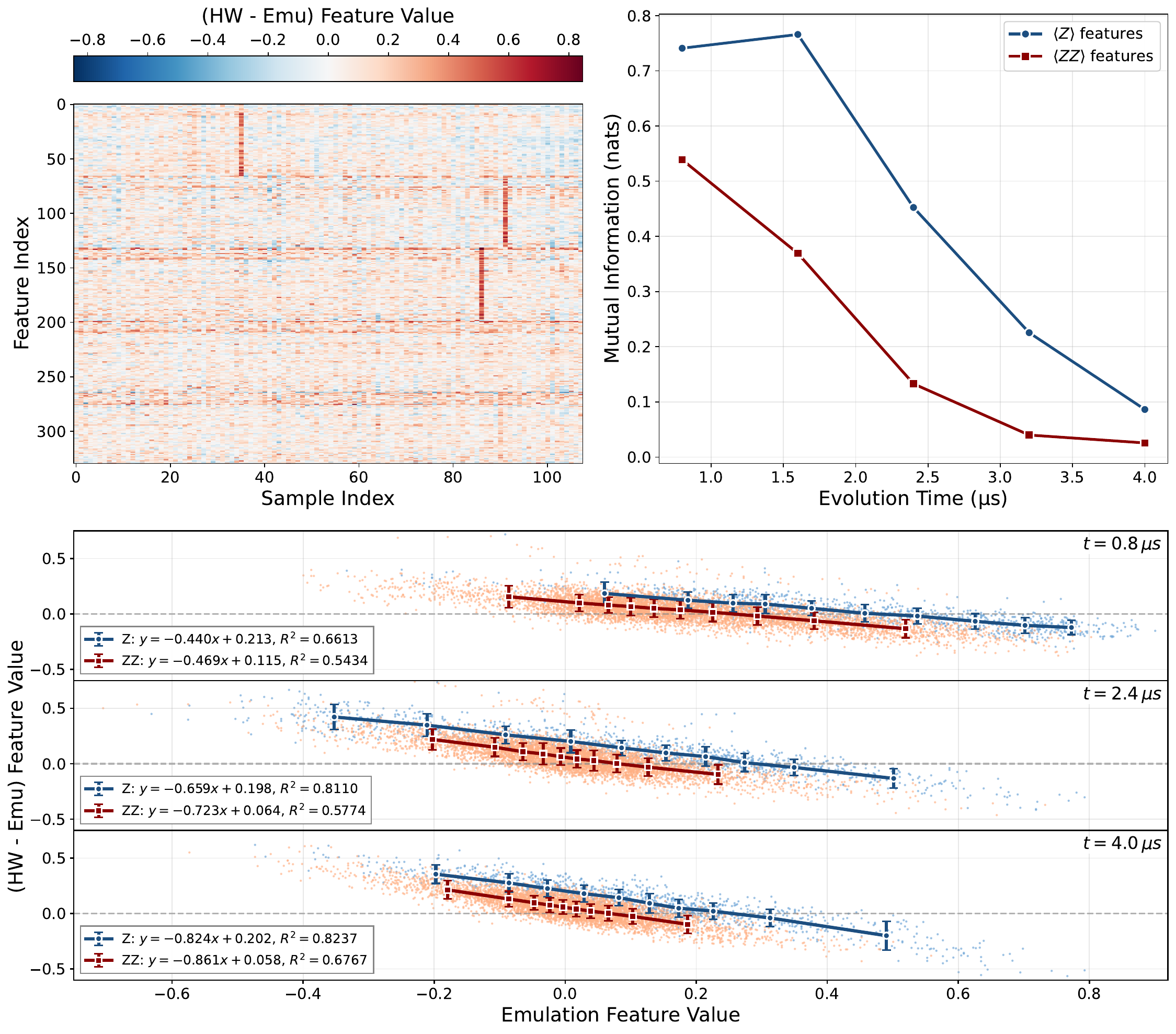}
\caption{\justifying
Comparison of hardware and emulated features across evolution times.
Top left: Heatmap of residuals ($\mathrm{HW}-\mathrm{Emu}$), showing structured sample- and feature-dependent deviations.
Top right: Mutual information between hardware and emulated features, decreasing at later times.
Bottom row: Residuals plotted against emulated values, with $\langle Z \rangle$ and $\langle ZZ \rangle$ features shown separately. Overlaid points indicate equal-population bins (0.01–0.99 quantile range), showing bin means and variances. Residuals follow an approximately linear trend with Gaussian-distributed fluctuations within each bin.
}
    \label{fig:hwemu_comparison}
\end{figure*}

To understand how hardware execution modifies the quantum features relative to noiseless emulation, we examine the relationship between hardware features ($\mathrm{HW}$) and their emulated counterparts ($\mathrm{Emu}$) across all samples and evolution times. Figure~\ref{fig:hwemu_comparison} summarises three complementary diagnostics.

We begin by visualising the residuals $(\mathrm{HW} - \mathrm{Emu})$ across all samples and features (Fig.~\ref{fig:hwemu_comparison}, top left), with samples ordered along the horizontal axis and features ordered along the vertical axis. If hardware deviations were dominated by unstructured stochastic noise, the residuals would appear randomly distributed. Instead, the residual heatmap exhibits clear structure. Vertical bands indicate sample-specific structure, where certain input samples systematically produce larger discrepancies, often confined to a single time step. Horizontal bands indicate feature-specific structure, with groups of features, particularly the single-site $\langle Z \rangle$ features, exhibiting consistently biased residuals across many samples. The presence of both sample- and feature-dependent patterns indicates that hardware deviations from emulation are structured rather than purely random, motivating a more detailed analysis of how hardware features depend on their emulated values.

To quantify this relationship, we plot the residuals $(\mathrm{HW} - \mathrm{Emu})$ as a function of the corresponding emulated feature values for representative evolution times (Fig.~\ref{fig:hwemu_comparison}, bottom). $\langle Z \rangle$ and $\langle ZZ \rangle$ features are shown in different colours, and a linear regression is performed, with fitted parameters and coefficient of determination ($R^2$) reported in the legend.

For both $\langle Z \rangle$ and $\langle ZZ \rangle$ features, the residuals exhibit an approximately linear dependence on the emulated values, with a negative slope indicating that hardware features are systematically shifted toward zero relative to emulation. The magnitude of this slope increases with evolution time, consistent with progressively stronger compression of feature values toward the mean. The linear fit provides a good description of this behaviour, with $R^2$ increasing at later evolution times, indicating that a larger fraction of the residual variation is captured by this linear relationship. The $\langle Z \rangle$ features additionally exhibit a positive intercept, corresponding to a constant upward shift of hardware values relative to emulation, likely caused by readout errors reported for \textit{Aquila}~\cite{wurtz_aquila_2023}.

To further characterise the residual statistics, we restrict the emulated feature range to the central quantiles (0.01–0.99) and partition the data into ten equal-population bins. For each bin, we compute and plot the mean residual and its variance. The binned means closely follow the linear trend obtained from the regression on the full dataset, confirming that the contractive relationship is not driven by outliers but is present across the feature range. Within each bin, the residuals are approximately Gaussian-distributed, with variance that remains relatively consistent across the feature range. This indicates that, in addition to the systematic contractive trend, a stochastic component contributes fluctuations around the mean transformation.

These observations provide insight into the improved downstream machine-learning performance observed when using hardware-generated features. In particular, the contractive mapping reduces the range of feature values, suppressing extremes and concentrating features closer to the mean. Such compression can reduce sensitivity to small input variations and limit the influence of outlier feature values. The approximately Gaussian residual fluctuations further introduce controlled variability around this mapping, reducing sensitivity to fine-scale structure present in the noiseless emulation.

Finally, to assess whether hardware execution preserves or alters the information content of the quantum features, we compute the mutual information
\begin{equation}
I(\mathrm{HW}_t;\mathrm{Emu}_t),
\end{equation}
between hardware and emulated features at each evolution time $t$ (Fig.~\ref{fig:hwemu_comparison}, top right). Mutual information quantifies the statistical dependence between the two distributions and provides a model-independent measure of how much information about the emulated features is retained in the hardware features. Mutual information is initially high at early evolution times, indicating a strong correspondence between hardware and emulated features, but decreases steadily with increasing evolution time. This reduction indicates that hardware execution progressively reshapes the feature representation rather than preserving a deterministic correspondence with emulation.

Taken together, the residual diagnostics and mutual information analysis show that hardware execution produces a feature representation that departs systematically from noiseless emulation in both its statistical structure and information content. Rather than acting as purely degradative noise, the hardware reservoir implements an effective stochastic feature map arising from the underlying physical dynamics and measurement processes. This transformation reshapes the geometry and statistical dependencies of the feature space, providing a concrete basis for the observed differences in downstream machine-learning performance and consistent with a hardware-induced modification that stabilises learning.

\section{Conclusion}
\label{conclusion}

In this work, we investigated the use of quantum reservoir computing (QRC) as a data transformation for downstream classical machine-learning on a biomarker classification task, with particular attention paid to differences between noiseless emulation and hardware execution.

We first established classical baselines using six models trained on the full dataset and subsequently on SHAP-ranked biomarker subsets. The majority of models trained on ranked subsets outperformed those using the full feature set, confirming that feature selection was beneficial for this dataset and providing a well-motivated input representation for subsequent quantum encoding.

Encoding these subsets into an emulated quantum reservoir produced quantum feature representations that achieved mean test accuracies comparable to their classical counterparts. However, models trained on emulated quantum features typically exhibited higher training accuracy and increased split-to-split variability, consistent with overfitting.

We then evaluated a hardware implementation of QRC using the neutral-atom processor \textit{Aquila}. Under the experimental constraint that biomarker subsets were aggregated across splits due to limited shot availability, we compared hardware features against emulated features using identical inputs. Across multiple models and feature-count configurations, hardware features yielded more robust performance across data splits and, in many cases, higher mean test accuracy. Standardised difference analysis confirmed that these effects were statistically significant.

To investigate the origin of this discrepancy, we analysed the statistical differences between hardware- and emulation-generated quantum features, which consist of expectation values of measured observables. Hardware execution was found to induce a structured, time-dependent transformation of these feature distributions, characterised by compression toward the mean and a reduction in mutual information between hardware and emulated features with increasing evolution time. These findings indicate that hardware dynamics systematically reshape the statistical structure of the feature representation, rather than acting as unstructured noise.

Overall, our results show that, for this dataset, noiseless quantum-reservoir feature generation does not outperform carefully constructed classical baselines. However, hardware execution introduces modifications to the feature representation that are consistent with a regularising effect and that frequently lead to improved robustness or accuracy relative to emulation. A definitive benchmarking of hardware QRC against classical approaches remains an open task, as shot limitations restricted the scope of the present study.

\section*{Acknowledgements}
This work was supported by resources provided by the Pawsey Supercomputing Research Centre's Setonix Supercomputer (https://doi.org/10.48569/18sb-8s43), with funding from the Australian Government and the Government of Western Australia. The Pawsey Supercomputing Research Centre's Quantum Supercomputing Innovation Hub and this work was made possible by a grant from the Australian Government through the National Collaborative Research Infrastructure Strategy (NCRIS). Access to the QuEra device Aquila was provided by the Pawsey Supercomputing Research Centre. LA acknowledges support from the Australian Government Research Training Program scholarships. LA, PJE and CRM thank Milan Kornjaca and John Tanner for helpful discussions. LA, PJE and CRM thank the entire Quantum Supercomputing Innovation Hub team at Pawsey, QuEra, and the little atoms used in this work.

\bibliographystyle{unsrt}
\bibliography{refs}

\clearpage

\appendix
\onecolumngrid
\section*{Appendices}
\addcontentsline{toc}{section}{Appendices}

\section{Hyperparameter Distributions for Random Search}
\label{app:hyperparams}

For each stage of model optimisation, hyperparameters were randomly sampled from predefined probability distributions. Continuous parameters were drawn from uniform ranges, while integer-valued parameters were sampled from discrete uniform distributions. Categorical hyperparameters were selected uniformly from finite sets. Table~\ref{tab:hyperparam_distributions} summarises the distributions used for each model.

\begin{table*}[th] \centering \footnotesize \caption{\justifying Hyperparameter distributions used for random search. \texttt{randint(a,b)} samples integers in $[a,b]$ and \texttt{uniform(a,b)} samples real values in $[a, a+b]$. 1000 combinations were sampled at all stages of hyperparameter optimisation.} \label{tab:hyperparam_distributions} \begin{tabular}{p{0.25\textwidth}|p{0.70\textwidth}} \toprule \textbf{Model} & \textbf{Hyperparameter Distributions} \\ \midrule \textbf{XGBoost} & \texttt{n\_estimators}: \texttt{randint(50, 200)} \newline \texttt{learning\_rate}: \texttt{uniform(0.01, 0.3)} \newline \texttt{max\_depth}: \texttt{randint(3, 10)} \newline \texttt{min\_child\_weight}: \texttt{randint(1, 10)} \newline \texttt{subsample}: \texttt{uniform(0.6, 0.4)} \newline \texttt{colsample\_bytree}: \texttt{uniform(0.6, 0.4)} \newline \texttt{reg\_alpha}: \texttt{uniform(0, 1)} \newline \texttt{reg\_lambda}: \texttt{uniform(1, 3)} \\ \midrule \textbf{Random Forest} & \texttt{n\_estimators}: \texttt{randint(50, 200)} \newline \texttt{max\_depth}: $\{3, 5, 7, 10, 15, \text{None}\}$ \newline \texttt{min\_samples\_split}: \texttt{randint(2, 20)} \newline \texttt{min\_samples\_leaf}: \texttt{randint(1, 10)} \newline \texttt{max\_features}: $\{\text{sqrt},\, \text{log2},\, \text{None}\}$ \newline \texttt{bootstrap}: $\{\text{True},\, \text{False}\}$ \\ \midrule \textbf{Gradient Boosting} & \texttt{n\_estimators}: \texttt{randint(50, 200)} \newline \texttt{learning\_rate}: \texttt{uniform(0.01, 0.3)} \newline \texttt{max\_depth}: \texttt{randint(3, 8)} \newline \texttt{min\_samples\_split}: \texttt{randint(2, 20)} \newline \texttt{min\_samples\_leaf}: \texttt{randint(1, 10)} \newline \texttt{subsample}: \texttt{uniform(0.6, 0.4)} \\ \midrule \textbf{Logistic Regression} & \texttt{C}: \texttt{uniform(0.01, 100)} \newline \texttt{penalty}: $\{\text{l1},\, \text{l2}\}$ \newline \texttt{solver}: $\{\text{liblinear},\, \text{saga}\}$ \\ \midrule \textbf{Support Vector Machine (SVM)} & \texttt{C}: \texttt{uniform(0.1, 100)} \newline \texttt{kernel}: $\{\text{linear},\, \text{rbf}\}$ \newline \texttt{gamma}: $\{\text{scale},\, \text{auto},\, \text{float}\}$ \\ \midrule \textbf{k-Nearest Neighbours (kNN)} & \texttt{n\_neighbors}: \texttt{randint(3, 20)} \newline \texttt{weights}: $\{\text{uniform},\, \text{distance}\}$ \newline \texttt{algorithm}: $\{\text{auto},\, \text{ball\_tree},\, \text{kd\_tree},\, \text{brute}\}$ \newline \texttt{p}: $\{1, 2\}$ \\ \bottomrule \end{tabular} \end{table*}

\section{Initial SHAP Analysis using Full Feature Set}
\label{app:initial_shap}

\begin{figure*}[ht]
    \centering
    \includegraphics[width=1\linewidth]{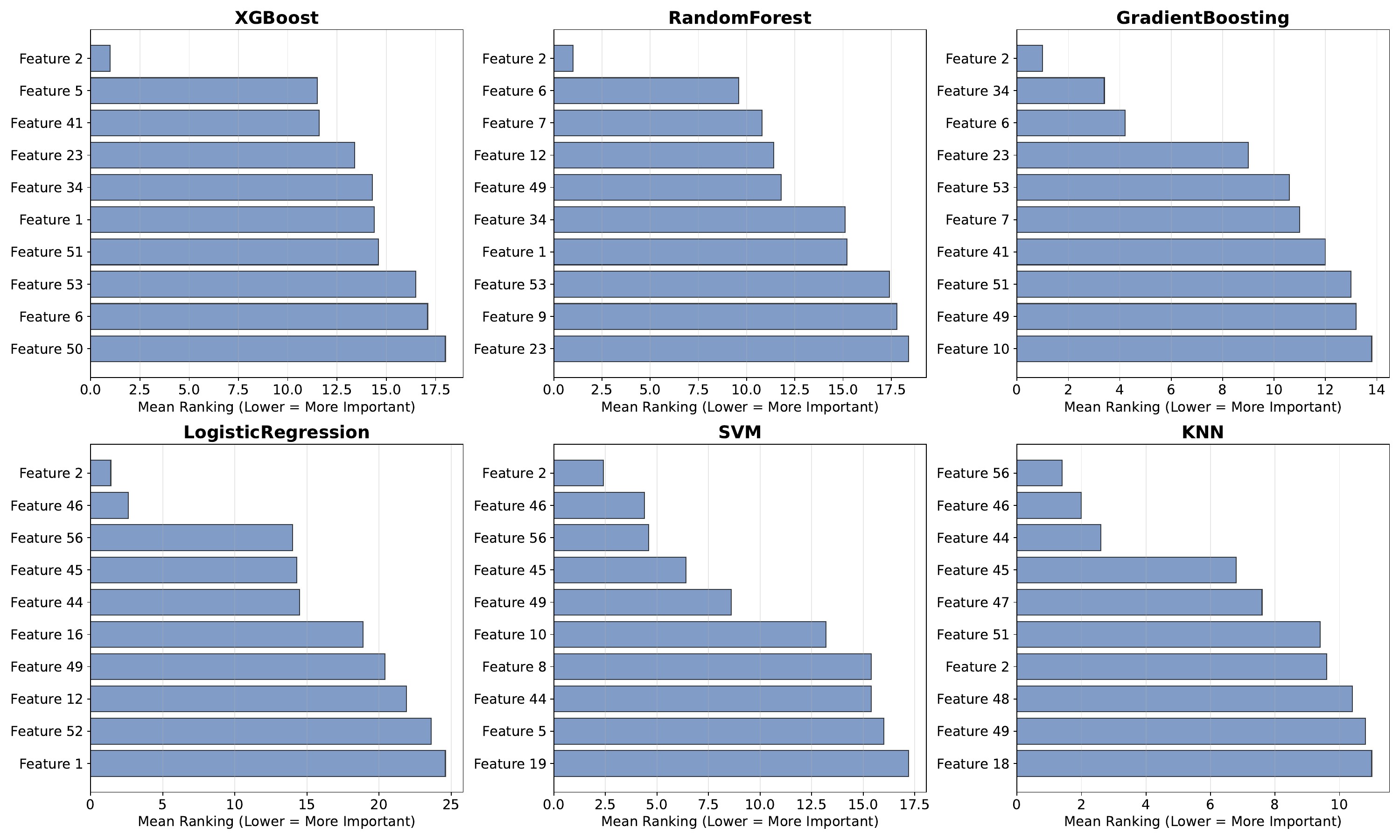}
    \caption{\justifying Mean SHAP ranking of the ten most important clinical features for each machine-learning model when trained using all data features. Each barplot reflects ranking averaged across all outer data splits. The six models naturally separate into two groups: the three tree-based models show strong agreement on one core subset of high-importance biomarkers, while Logistic Regression, SVM, and kNN agree on a different core subset. \texttt{Feature 2} appears in all six models.}
    \label{fig:shap_top10}
\end{figure*}

To characterise how different classical models prioritise clinical biomarkers, for each outer split of the data we computed the absolute SHAP value of each feature. To do this we used the optimised models obtained during the initial hyperparameter search with the full feature set. Figure~\ref{fig:shap_top10} shows the ten most influential features identified by each machine-learning algorithm. The features are ordered by their mean ranking across all outer splits. While there is broad agreement on a core subset of high-importance biomarkers, the models also exhibit notable differences in how they weight secondary features, reflecting their distinct inductive biases and interaction structures. These patterns help motivate the aggregated SHAP feature subsets used for the QRC hardware experiments, as several models assign similar rankings to the top features despite architectural differences.

\section{Model Performance with Quantum Features from Hardware}
\label{appendix:hardware_aggregate}

In this appendix we present model performance using quantum features generated on hardware under a modified feature-selection procedure. Unlike the results shown in Sec.~\ref{model_performance}, where SHAP-ranked feature subsets were determined independently for each data split, here we construct a single aggregated SHAP ranking across all splits and use this fixed feature ordering to encode data for both hardware and emulation. This aggregation was required to minimise the number of hardware shots required.

\begin{figure*}
    \centering
    \includegraphics[width=\linewidth]{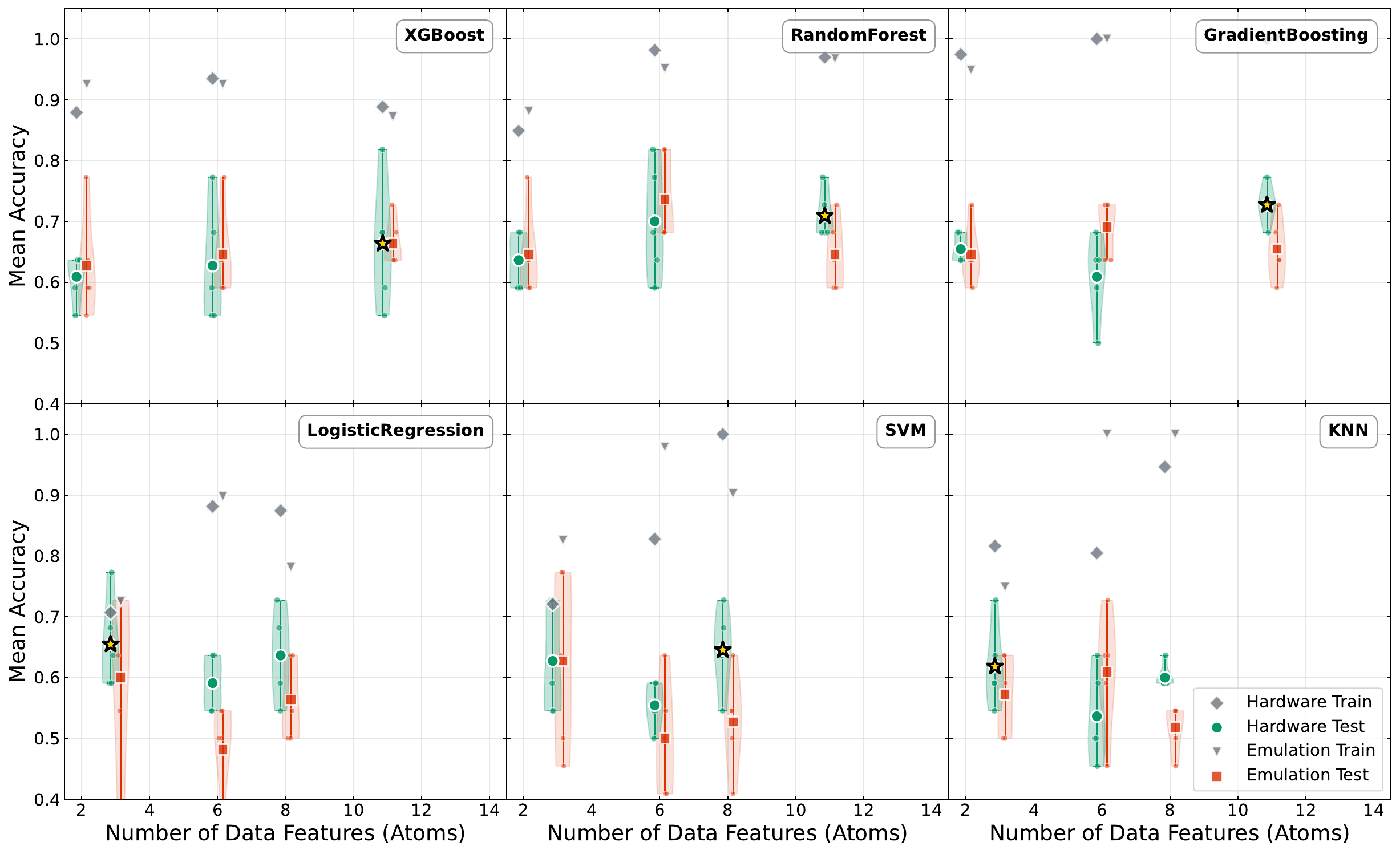}
    \caption{\justifying
    Model performance using hardware and emulated quantum features generated  with aggregated SHAP feature encoding. Mean train and test accuracy are shown across feature counts, with violin plots indicating variability over data splits. Because the feature ranking is computed using information aggregated across all splits, these results are not directly comparable to the classical or emulated performance reported in the main text and should not be interpreted as unbiased performance estimates. Instead, the figure enables a controlled comparison between hardware and emulation under identical feature encodings.}
    \label{fig:hardware_violins}
\end{figure*}

Because the aggregated ranking incorporates information from all data partitions, it introduces potential information leakage between training and test sets. Consequently, the resulting performance metrics cannot be interpreted as unbiased estimates of real-world predictive performance. For this reason, the results in Fig.~\ref{fig:hardware_violins} should not be compared directly to those reported in Figs.~\ref{fig:classical_performance} and~\ref{fig:qrc_emulated_performance}. For example, using this aggregated feature set we notice that the variability across splits for emulated QRC is lower than see in the main text. An analysis of whether this is caused by data leakage or whether this aggregate feature ordering is more suitable for QRC is deferred to future work.

Instead, this figure serves a narrower purpose: it provides a controlled comparison between hardware and emulated quantum features when both are constructed using the same aggregated encoding. Under these matched conditions, differences in model performance can be attributed to the feature-generation process rather than to variations in feature selection.

In most cases the hardware results have higher mean test accuracy than from emulation. In several configurations the results from hardware also have a lower variability across different splits of the data. These observations are discussed more in Sec.~\ref{hardware_results}.

\end{document}